# Improved 3-Dimensional Security in Cloud Computing

## Sagar Tirodkar[1], Yazad Baldawala[1], Sagar Ulane[1], Ashok Jori[1]

[1](Department of Computer Engineering, Dr. D. Y. Patil College of Engineering, Ambi/University of Pune, India)

*Abstract: Cloud computing is a trending technology in the field of Information Technology as it allows sharing of resources over a network. The reason Cloud computing gained traction so rapidly was because of its performance, availability and low cost among other features. Besides these features, companies are still refraining from binding their business with cloud computing due to the fear of data leakage.*

*The focus of this paper is on the problem of data leakage. It proposes a framework which works in two phases. The first phase consists of data encryption and classification which is performed before storing the data. In this phase, the client may want to encrypt his data prior to uploading. After encryption, data is classified using three parameters namely Confidentiality [C], Integrity [I] and Availability [A]. With the help of proposed algorithm, criticality rating (Cr) of the data is calculated. According to the Cr, security will be provided on the basis of the 3 Dimensions proposed in this paper.*

*The second phase consists of data retrieval by the client. As per the concept of 3D, users who want to access their data need to be authenticated, to avoid data from being compromised. Before every access to data, the user's identity is verified for authorization. After the user is authorized for data access, if the data is encrypted, the user can decrypt the same.*

*Keywords— Availability, Cloud Security, Confidentiality, Data Leakage, Decryption, Encryption, Integrity*

## I. INTRODUCTION

Cloud computing is nothing but a specific style of computing where everything from computing power to infrastructure, business apps etc. are provided "as a service". Cloud computing generally works on three types of architectures. These are: SAAS, PAAS and IAAS [1].

- Software as a Service (SAAS) – Users are provided access to application software and database. Cloud vendors manage the platforms and infrastructure that run the applications.

- Platform as a service (PAAS) – Cloud vendors provide a platform for computing, which includes the OS, coding language, execution environment, web server and the database. Developers can use the resources to build and run their software without buying expensive hardware.

- Infrastructure as a service (IAAS) – Providers of IAAS offer computers – physical or virtual – and other resources. A good example of IAAS is dedicated servers provided by Web Hosting sites such as Bluehost, Hostgator etc.

Cloud computing is making everything simpler and flexible nowadays, but there is another important aspect which is "What about the security of data over the cloud?" Cloud architecture with robust security implementation is the key to cloud security. Cloud is complex and hence security measures are not simple too. Since it is new, it faces new security issues and challenges as well. Till date, most users don't trust storing their data on SASS-based cloud computing providers such as Dropbox, Skydrive and Google Drive etc. Since the outburst of Cloud Computing in the year 2006, various methods are devised to increase the security of the data being stored over Cloud Servers. Some of which include Encryption, Decryption, Data Partitioning, Digital Signatures etc.

## II. LITERATURE REVIEW

From its inception, in the last few years cloud computing has emerged from just being a concept to constitute a major part of the IT industry.

To find out the top threats, CSA conducted a survey of industry experts to compile professional opinion on the greatest weaknesses within cloud computing [2]. In a recent edition report, experts identified nine threats which are critical to cloud security (ranked in order of severity):

TABLE I
THREATS TO CLOUD SECURITY

| Threat Rank | Threat | Is The Threat Still Relevant (%) | | |
|---|---|---|---|---|
| | | Yes | No | Needs Update |
| 1 | Data Breaches | 91 | 4.5 | 4.5 |
| 2 | Data Leakage | 91 | 4.5 | 4.5 |
| 3 | Account Hijacking | 87 | 9 | 4 |
| 4 | Insecure APIs | 90 | 7 | 3 |
| 5 | Denial of Service | 81 | 16 | 3 |
| 6 | Malicious Insiders | 88 | 8 | 4 |
| 7 | Abuse of Cloud Services | 84 | 14 | 2 |
| 8 | Insufficient Due Diligence | 81 | 16 | 3 |
| 9 | Shared Technology Issues | 82 | 14 | 4 |

Gartner surveyed more than 300 cloud users, verifying the top three concerns among them [3]. Nearly 50% of the response identified service provider security as their primary problem. Tier1 Research's 2011 report on the hosting market indicates that most of the enterprises consider securing infrastructure as the most problematic aspect of the cloud. In a report based of 1500 users by the Alert Logic Security research team, security incidents identified through a combination of automated correlation and validation by security analyst. According to this report, when compared to traditional in-house privates cloud environment, cloud providers show lower occurrence rates for every class of incident examined. Service provider customers experienced lower threat diversity than on-premise customers. On-premise cloud systems are twelve times more likely than service provider environments to have common configuration issues, allowing the data to be compromised. While conventional wisdom suggests a higher rate of Web application attacks in the service provider environment, we found a higher frequency of these incidents in on-premise environments.

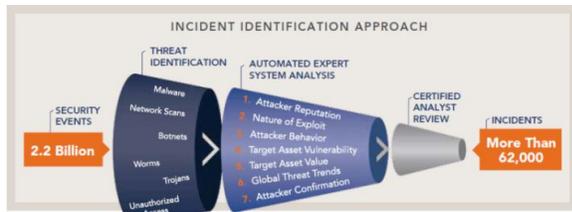

Fig 1: Incident Identification Approach

As shown in Fig. 1, 2.2 Billion Security events observed during the study period were automatically evaluated and correlated through Alert Logic's expert system and reviewed by security analysts. More than 62,000 incidents were verified and classified into seven incident categories.

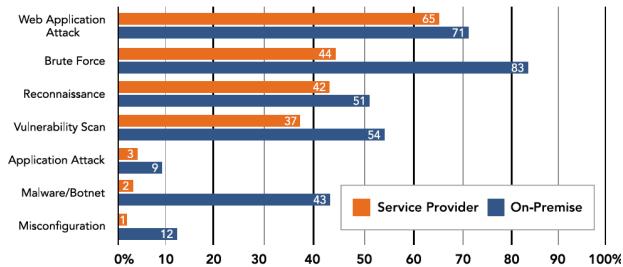

Fig 2: Percentage of Cloud Customers experiencing Security Incidents

As seen in Fig. 2, the rate of occurrence in an on-premise environment is more likely to be greater than the occurrence rate for service provider customers. This observation is true for all threat categories. The frequency of experienced incidents is higher for on-premise environments across most of the threat categories. The threat diversity for on-premise environments is greater than the threat diversity for service provider environments.

### 1. Related Work Done

There is a critical need to securely store, manage, share and analyse data to improve the security of cloud data. The major security challenge with data uploaded over the cloud is that the owner of the data might not have sufficient control over it.

Therefore we need to safeguard the cloud ecosystem in the midst of un-trusted forces. Some of the work done in trying to achieve the said objective is as follows:

**1.1 Encrypted Data Storage for Cloud Computing:** Since the data uploaded over the users on the cloud is placed on remote servers, it may be necessary that the data is encrypted. Encryption adds an additional layer of security to the data. Encryption can be implemented by various algorithms, some of which include RSA, DES, and SHA [4],[5]. However, one of the disadvantages of encryption is that if the Encryption Keys are managed by any other entity e.g. The Cloud Provider, then the data is still not as safe as it needs to be.

**1.2 Data Partitioning Technique:** This method uses a third party auditor for facilitating the data manipulation [6]. One of the advantages of this system is that even if data leakage takes place, the data cannot be accessed as it is a cluster of parts of the whole data.

This literature survey was initiated to better understand the particular reasons that the users feel sceptical towards cloud computing. We also wanted to explore different ways through which we can alleviate these concerns and boost the users' confidence in cloud computing. Single provider in the IT ecosystem: hardware, software, or Cloud service provider cannot single-handedly address IT's cloud security issues. Various researches show that IT professionals have serious concerns about control over data and access.

Solving these issues will require efforts across the Information Technology industry. Industry-wide security standards and common frameworks that enable IT professionals to exercise restraint and measure their cloud environments will aid in ensuring the use of cloud computing worldwide. If the IT industry works together to advance cloud security standards and capabilities, we believe that cloud computing has the potential to offer the same or even greater security and safety than is provided by the traditional IT infrastructures of today.

### III.    PROBLEM STATEMENT

The focus of this paper is on the problem of data leakage. It proposes a framework which works in two phases. The first phase consists of data encryption and classification which is performed by the client before storing the data. In this phase, the client may want to encrypt his data prior to uploading. DES algorithm is used for encryption/decryption of user data. After encryption data is classified using three parameters namely Confidentiality [C], Integrity [I] and Availability [A]. The client needs to specify the values of these parameters while storing the data. The value of Confidentiality [C] is based on the level of security required for the data. The value of Integrity [I] provides the scale to which the accuracy and reliability of the information stored is required. The value of Availability [A] is based on the frequency at which the data will be accessed. With the help of proposed algorithm, criticality rating of the data is calculated. According to the criticality rating of the data, security will be provided on the basis of the 3 Dimensions proposed in this paper [7].

After the completion of first phase, the second phase consists of data retrieval by the client. As per the concept of 3D, users who want to access their data need to be authenticated, to avoid data from being compromised. Before every access to data, the user's identity is verified for authorization. The Outermost Ring comprises of having the user enter the password to access user data. The Mid Ring consists of Authentication with Graphical password[8]. The Innermost Ring consists of the most secure data and hence consists of Authentication with One-Time Password[9]. After the user is authorized for data access, if the data is encrypted, the user can decrypt the same using DES Algorithm.

## 1. Methodology

In this technique, the first job of the user is to provide data for storage over the cloud. In this the user has the option to Encrypt the Data if he wishes to before uploading the same. This adds another layer of security to the users' data. After this point the user can upload the Encrypted data to the Cloud. While doing the same, the user provides the value of Confidentiality [C], Integrity [I] and Availability [A]. After applying the proposed algorithm for data classification, a criticality rating [Cr] is calculated for Data D[]. Now allocation of data on the basis of Cr is done in a protection ring. This technique suggests that the innermost protection ring is very critical and it requires a robust security technique to ensure high confidentiality for the data.

## 2. Phase I: Data Encryption

Prior to uploading the Data users have an option to Encrypt the Data. This allows the user to secure his data even further.

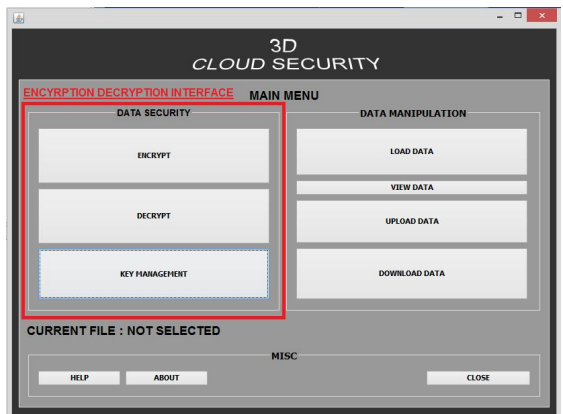

Fig 3: Encryption/Decryption Interface

One of the major concerns of Encryption is that "Who manages the Encryption Keys?" For Security Reasons it is best that the Encryption Keys are managed by the users as shown in Fig 4 below.

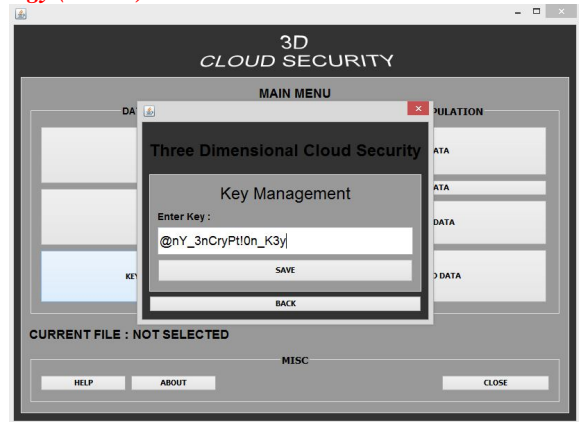

Fig 4: Key Management

## 3. Phase I: Data Classification

1. Input: User provides the data as input along with CIA parameters of security levels. User provides the value of CIA parameters. User place, protection ring, arrays D, C, I, A, S, R of n integer size.

2. Output: Data is classified into three protection rings i.e. ring 1, ring2, and ring 3. This aids in providing different levels of access to different categories of data. Define data protection ring D[] array of n size.

3. For i = 1 to n //Input values of CIA
   C[i] – value for i'th data
   I[i] – value for i'th data
   A[i] – value for i'th data

4. For i=1 to n
   /*assume rings 3 and protection levels from 1 to 10 selected manually by client while uploading files*/

   If (C[i]>6 and I[i]>6) then
   Ring level=1 (high)
   Else
   Calculate new average term CI of both C & I
   CI=(C[i] + I[i])/2
   Go to step 5

5. For i=1 to n
   If (CI>3 and CI<5 and A[i] <5)
   Then Ring level 2(Mid)
   If (CI>3 and CI<5 and A[i]>5) then
   Ring level 3(Low)
   If (CI>=1 and CI<=3)
   Then Ring level 3(Low)

In above algorithm the first job of the user is to upload the data. The proposed framework uses data classification algorithm to classify data on the basis of Confidentiality, Integrity and Availability. Here D[] represents data. The user has to give the value of C-Confidentiality, I-Integrity and A-Availability. After applying proposed formula the value of Criticality Rating (Cr) is calculated. Data is allocated to a protection ring based on the Cr. This technique suggests that internal protection ring is very critical and it requires more security technique to ensure confidentiality.

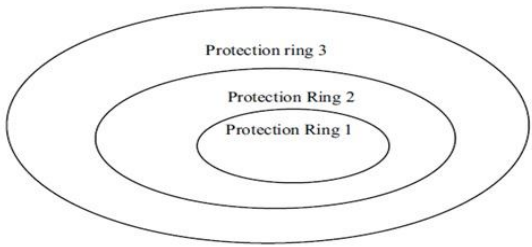

Fig 5: Protection Rings Setup

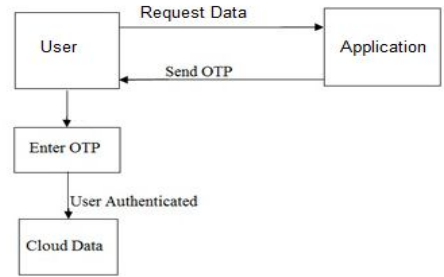

Fig 7: One-Time Password Methodology

After classification of data in above step, three entities are considered, first one is cloud provider itself, second is organization whose data resides at cloud and last one is user who request for access of cloud data. As seen above, in the first phase Data is classified according to its criticality into Ring 1, Ring 2, and Ring 3.

### 4. Phase II: Data Access

Now the Fig. below gives an overview of the first step of the second phase in which the Methodology for the Data Access is selected. Ring 1 has the most Critical Data and hence is given the highest level of Personal Access.

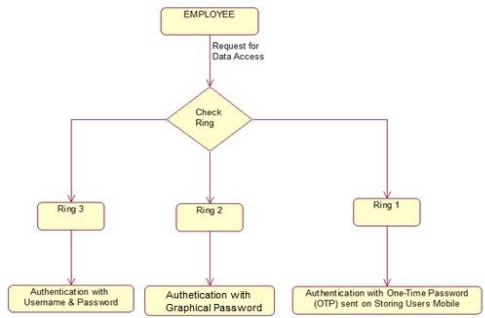

Fig 6: Data Access Methodology

Now if a user want to access the data, if it belongs to protection ring 1, it require authentication with OTP sent to Users Mobile, if the data belongs to ring 2 then user have to Authenticate with Graphical Password, if the data belongs to ring 3 then it just needs the user to re-enter the password. Now suppose the user registered itself for accessing data, the Client Software will provide Username, Password Multi-Dimensional Password for Authentication as well as facilitate the OTP Generation on the Users Mobile.

### 5. Ring 1: Authentication with One-Time Password Sent on Users Mobile

When user sends request along with username to access the data to cloud provider, the cloud provider first checks the user mobile number, it then generates the OTP and sends it to the Users Mobile.

Now the user needs to enter the OTP received for authentication, and after authentication access to the data will be provided.

*5.1 Algorithm: One-Time Password Generation*

i.  A Random String is generated using the user data as follows:

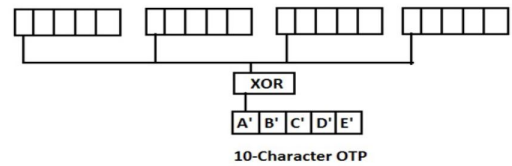

ii. This String is converted to 160-bit SHA-1 hash function.

iii. This 160-bit (20 bytes) Hash is split into 4 groups of 5 byte each.

iv. The 1st byte of each group is XORed to get 1st byte of OTP. Similarly the remaining bytes of OTP are generated, as shown below.

v.  The XORed String ABCDE is converted to Hexadecimal and Sentto User's Mobile as an OTP.

The Fig. 5 below shows the Interface to enter the OTP sent to the User's Mobile. Once the user enters the OTP it verifies it with the OTP generated and then provides access to the data. Generally the OTP should be valid for a limited period of time say 10 minutes.

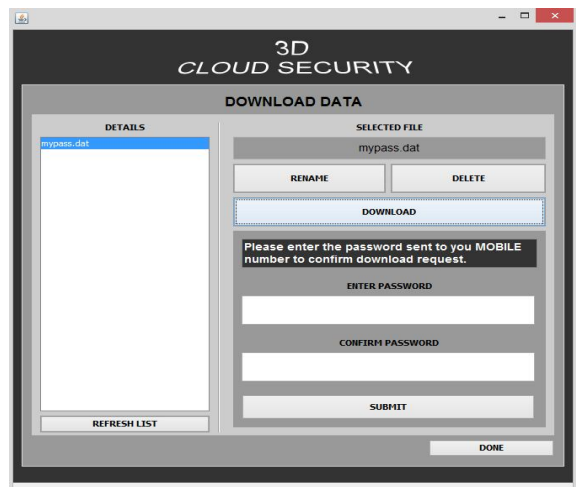

Fig 8: OTP Interface

## 6. Ring 2: Authentication with Graphical Password

According to this technique, access to the cloud is authenticated using a graphical password. The Graphical password is generated by considering many aspects and confidential inputs of images. With the help of this technique, the probability of brute force attack for breaking the password is greatly reduced.

### 6.1 Algorithm: Graphical Password Selection

1. Read Input values as Image IDs.
2. Group Image IDs to form the password.
3. Send Generated Password to Database.
4. Finish.

In this technique, the user is provided with three sets of eight Images and is required to select one image from each set as shown in Fig. 6.

The images contains a unique ID which is stored in the Database as set of 3 ID's to be used for Authentication. When the user needs to access his data, the three sets are displayed again for authentication. Also the position of the images in each set is always shuffled, in order to further reduce the probability of brute force attack.

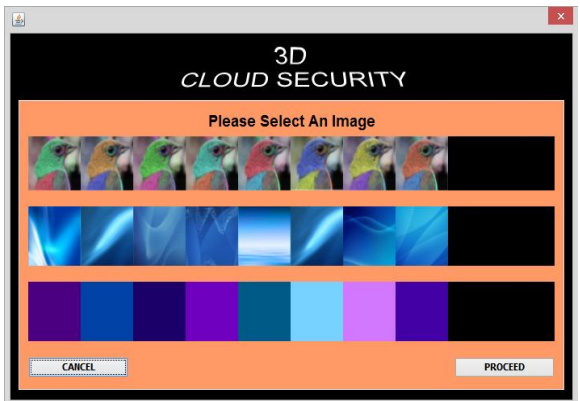
Fig 9: Graphical Password Selection Interface

## 7. Ring 3: Authentication with User Password

When user sends request to download data from the cloud provider, if the data is in Ring 3, the Cloud Provider asks the user to re-enter the Password for Authentication. After the Authentication it allows the User to access the Resource.

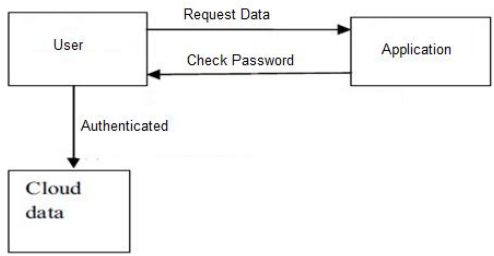
Fig. 10: Password based Authentication Methodology

## 8. Phase II: Decryption

Decryption Allows user to decrypt the Data after it has been downloaded. This allows only the User to access to the data as he is the only one who has access to the Key used for Encryption. We have showed the Decryption and Key Management interface in Section III.B Fig.3. Once the data is decrypted the user will be able to view it normally.

## IV. CONCLUSION & FUTURE WORK

This technique provides a new way to authenticate in 3 Dimensional approaches as well as providing an optional Encryption Framework. The data availability is provided by overcoming many existing problems like Denial of Services, data leakage, user managed encryption keys etc. It also provides more flexibility and capability to overcome the problems faced by today's complex and diverse networks.

### 1. Future Work

Customers usually raise data security concerns related to:
- Better Security
- Regulatory Compliance
- Flexible deployment provisioning
- Dealing with Complexity
- Key Management in case of encryption

There is a strong industry agreement that security, parallel with regulatory compliance, is the first obstacle to adoption of cloud computing [10]. Underlining these concerns is the need to establish trust an organization can outsource its storage or its compute resources, but confidentiality cannot be outsourced.

At the same time, companies are attracted to cloud computing for its various advantages such as flexibility, elasticity and the easy economic model. Cloud customers can bring up servers and storage in short time and they expect a safety solution to provide the same high level of automation and management.

Customers cannot accept a trade-off between security and flexibility. The customer expects the cloud vendor to deliver the best security of data which does not compromise the cloud values of flexibility and elasticity. This may be a tedious task.

A solution is required for these factors: up in minutes; pay as you go; using the strongest proven encryption algorithms; and ensuring verification of records and adherence to laws.

The cloud vendor must keep the customer's data always encrypted, and the encryption keys should themselves be encrypted, even when in use. The technologies such as Key splitting and homo-morphing will aid in solving security challenges. All these combined technologies will provide a breakthrough in the field of Information Technology.

### REFERENCES

[1] Meiko Jensen etal, "On technical security issues in cloud computing", *CLOUD '09. IEEE International Conference on Cloud Computing, 2009, 109-116*

[2] Cloud Security Alliance "The Notorious Nine: Cloud Computing Top Threats in 2013".[Online]. Available: https://cloudsecurityalliance.org/research/top-threats/

[3]   Alertlogic "State of Cloud Security Report | spring 2012". [Online].
      Available: https://alertlogic.com/resources/cloud-security-report/

[4]   S. Han, H. Soo, "The Improved data encryption standard (DES)
      algorithm", Spread Spectrum Techniques and Applications
      Proceedings, 1996., *IEEE 4th International Symposium on
      (Volume:3), Sep. 1996*

[5]   A. Chhabra, "Modified RSA Algorithm: A secure Approach", *2011
      International Conference on Computational Intelligence and
      Communication Networks (CICN), 2011, 545-548*

[6]   C. Selvakumar, "Improving cloud data storage security using data
      partioning technique", *IEEE 3rd International Advance Computing
      Conference (IACC), 2013, 7-11*

[7]   P. Prasad, B. Ojha, R. Shahi, R. Lal, A. Vaish, U. Goel, "3
      Dimensional Security in Cloud Computing", *3rd International
      Conference on Computer Research and Development (ICCRD), 2011,
      198-201*

[8]   A.M. Eljetlawi, "Graphical password : Existing recognition base
      graphical usability", *6th International Conference on Networked
      Computing (INC), 2010, 1-5*

[9]   S. Acharya, A. Polawar, P. Pawar, "Two factor authentication using
      smartphone generated one time password", *IOSR Journal of Computer
      Engineer, Volume 11, 2013*

[10]  Info Security Products Guide. [Online]. Available at:
      http://www.porticor.com/2012/12/cloud-security-qa-gpn/